# On the Formation of Hydrogen Peroxide in Water Microdroplets

Adair Gallo Jr. [1,&], Nayara H. Musskopf[1,&], Xinlei Liu[2], Ziqiang Yang[2], Jeferson Petry[1], Peng Zhang[1], Sigurdur Thoroddsen[2], Hong Im[2], Himanshu Mishra[1,*]

[1]Interfacial Lab (iLab), Biological and Environmental Science and Engineering (BESE) Division, Water Desalination and Reuse Center (WDRC), King Abdullah University of Science and Technology (KAUST), Physical Science and Engineering (PSE), Thuwal 23955-6900, Saudi Arabia

[2]Physical Sciences and Engineering (PSE) Division, King Abdullah University of Science and Technology (KAUST), Physical Science and Engineering (PSE), Thuwal 23955-6900, Saudi Arabia

[&]Equal author contribution
*Himanshu.Mishra@kaust.edu.sa

**Abstract**

Recent reports on the formation of hydrogen peroxide ($H_2O_2$) in water microdroplets produced via pneumatic spraying or capillary condensation have garnered significant attention. How covalent bonds in water could break under such conditions challenges our textbook understanding of physical chemistry and the water substance. While there is no definitive answer, it has been speculated that ultrahigh electric fields at the air-water interface are responsible for this chemical transformation. Here, we resolve this mystery via a comprehensive experimental investigation of $H_2O_2$ formation in (i) water microdroplets sprayed over a range of liquid flow-rates, the (shearing) air flow rates, and the air composition (ii) water microdroplets condensed on hydrophobic substrates formed via hot water or humidifier under controlled air composition. Specifically, we assessed the contributions of the evaporative concentration and shock waves in sprays and the effects of trace $O_3(g)$ on the $H_2O_2$ formation. Glovebox experiments revealed that the $H_2O_2$ formation in water microdroplets was most sensitive to the air-borne ozone ($O_3$) concentration. In the absence of $O_3(g)$, we could not detect $H_2O_2(aq)$ in sprays or condensates (detection limit ≥250 nM). In contrast, microdroplets exposed to atmospherically relevant $O_3(g)$ concentration (10–100 ppb) formed 2–30 μM $H_2O_2(aq)$; increasing the gas–liquid surface area, mixing, and contact duration increased $H_2O_2(aq)$ concentration. Thus, the mystery is resolved – the water surface facilitates the $O_3(g)$ mass transfer, which is followed by the chemical transformation of $O_3(aq)$ into $H_2O_2(aq)$. These findings should also help us understand the implications of this chemistry in natural and applied contexts.

**Introduction**

Interfacial mass and energy transfer, chemical transformations/reactions, and tensions at water interfaces are implicated in numerous natural and industrial processes, such as the atmosphere–hydrosphere exchange[1], cloud physics[2] and chemistry[3], thundercloud charging[4, 5], precipitation[6], ion speciation[7] and ion-catalyzed proton transfers[8, 9], aerobic bioreactors[10], food science[11], soil physics[12], underwater insect respiration[13], and the ubiquitous water evaporation/condensation events. Probing the air–water interface of molecular thickness, however, remains a daunting research area due to the challenges associated with direct experimental interrogation as well as the limitations of the water models[14]. Therefore, investigating the properties of the water surface is a research frontier in chemical science, sometimes invoking vigorous debates[14-25]. With this preface, we introduce the latest reports on the chemical transformation of water into hydrogen peroxide ($H_2O_2$). Researchers have discovered the spontaneous production of (i) $H_2O_2$ (~30 μM) in water microdroplets of diameter ≤ 20 μm sprayed via pressurized gas[26] and (ii) $H_2O_2$ (≤ 115 μM) in condensed water microdroplets of diameter ≤ 10 μm on common substrates in the relative humidity range 40–70%[27]. Crucially, smaller droplets yielded higher $H_2O_2$ concentration, pinpointing that the air-water interface is somehow implicated. How could the O–H covalent bonds in water with an approximate strength of ~100 kcal-mol$^{-1}$ be broken under normal temperature and pressure (NTP, 293 K and 1 atm; $k_B T = 0.58$ kcal-mol$^{-1}$) without the use of a catalyst, significant energy, or co-solvents? There is no explanation available at the moment. Presence of ultrahigh electric fields at the air–water interface has been speculated to be the cause[26, 27]. These findings challenge our textbook-level understanding of water and perhaps other similar liquids where constituent atoms occupy distant positions on the Pauling's electronegativity table[28]. Given water's innumerable roles in environmental cycles, processes, and phenomena, could these findings herald a reassessment of what we know of water; opportunities in green chemistry also deserve a serious consideration.

We started exploring this problem soon after the appearance of the first report[26]. A follow-up report revealed that even condensed water microdroplets contained $H_2O_2$, in fact, at even higher concentrations ≤ 115 μM[27]. Here, we take a moment to juxtapose these two unrelated experimental methods to study this chemical transformation. Condensation is a gentle process governed by the relative humidity and the substrate's temperature and chemical make-up. In contrast, pneumatic sprays utilize high shearing gas-flow (e.g., 1–10 L/min) from a microscale capillary/annulus leading to fast gas speeds (e.g., 100–1000 ms$^{-1}$) that cause much turbulence/mixing/forced-convection inside and around the spray plume and can enhance the concentration of $H_2O_2$ due to the faster evaporation of the solvent (more details in Results and Discussion). In fact, if the spray-based experiments are conducted inside a controlled environment glovebox, the shearing gas-flow disturbs the atmospheric composition/distribution, especially the relative humidity. Thus, our preliminary investigation was restricted to water condensation[29], wherein we found that: (i) the condensed microdroplets formed via the vaporization of hot water (50–70°C) did not contain $H_2O_2$; (ii) if ultrasonic humidifiers were exploited to form the vapor, ~1 µM $H_2O_2$ was detected[29]. Therefore, we established that the ultrasonic humidifier was a contributor to the $H_2O_2$ production and the air–water did not have any detectable effect. The fact that ultrasonic waves can produce $H_2O_2$ in water is well-known and has even been exploited in practical applications, including disinfection[30] and water treatment[31] (see references[32-34] for further information). Still, however, we remained puzzled – *why did researchers in California observe ~115-times and ~30-times higher $H_2O_2$ in their condensation and spray experiments, respectively, while we did not find any $H_2O_2$ in similar experiments at KAUST?* Here, by developing and applying a broad set of experiments, we resolve this mystery for both sprayed and condensed microdroplets.

**Results**

In this experimental study, we utilized high-purity HPLC-grade water as well as water from the standard MilliQ Advantage 10 set up (Methods). Commercially available hydrogen peroxide ($H_2O_2$) 30% solutions (v/v) were diluted using water and utilized for the calibration of the assays and other experiments (Fig. S1). We utilized the fluorometric Hydrogen Peroxide Assay Kit (HPAK, ab138886, Abcam PLC, Cambridge, UK) that facilitated a detection range for aqueous $H_2O_2$ down to ~250 nM[35, 36], thereby affording a nearly 40-times lower limit of detection than the potassium titanium oxalate (PTO) assay employed in the original reports[26, 27] (Methods). HPAK calibration plots and comparisons with other standard $H_2O_2$ quantification methods, including the PTO assay and the hydroxyterephthalic acid (HTA) assay, have already been presented in our recent report[29].

For spraying water microdroplets, we custom-built pneumatic setups following the details presented in the recent report[26]. Broadly, our spray units comprised of concentric stainless steel tubes (Figs. 1 and S2), wherein water was pushed through the inner tube while pressurized $N_2$ gas was applied through the outer annulus to shear the liquid (Table S1). As water slugs/droplets moving at low speed are hit by fast flowing $N_2$ gas, microdroplets are produced; size distribution of microdroplets depends on the liquid and gas flow rates as well as capillary dimensions (Figs. S3-4). A sealed glass enclosure connected to the spray facilitated sample collection with minimal losses (Fig. 1). We also revisited the investigation of $H_2O_2$ concentration in condensed water microdroplets formed by heating water and ultrasonic humidifiers. Next, we delineate the various hypotheses and factors we considered could contribute to the $H_2O_2$ formation in water microdroplets (sprayed or condensed):

i. **Evaporative concentration (Figs. 1-2)**: trace amount of $H_2O_2$ could be present even in the water obtained from reverse osmosis[37]. During spraying, as the solvent evaporates, the solute might concentrate. NB: the boiling point of $H_2O_2$ at 1 atm is 423K that is 50 K higher than that of water[38, 39]; so $H_2O_2$ is expected to evaporate slower than water in the sprays and concentrate.

ii. **Mechanical vibrations, shock-waves, and cavitation in sprays (Fig. 3)**: As liquid flows through a capillary, then, depending on the liquid thermophysical properties, flow rate, shearing gas-flow rate, dissolved gases, capillary geometry, etc., shock waves and cavitation events can take place[40-42]. Cavitation implosion of bubbles in water can lead to extremely high temperatures and pressures in localized "hot spots", leading to the production of OH radicals that could yield $H_2O_2$[31].

iii. **Dissolution of airborne ozone in water and its autodissociation (Figs. 4–6)**: atmospheric/ambient ozone gas could dissolve in water and react to form $H_2O_2$[43]. N.B. the Henry's law constants for the solubility of ozone and $H_2O_2$ in water are, respectively, $\sim 10^{-3}$ M/atm and $\sim 10^4$ M/atm [44].

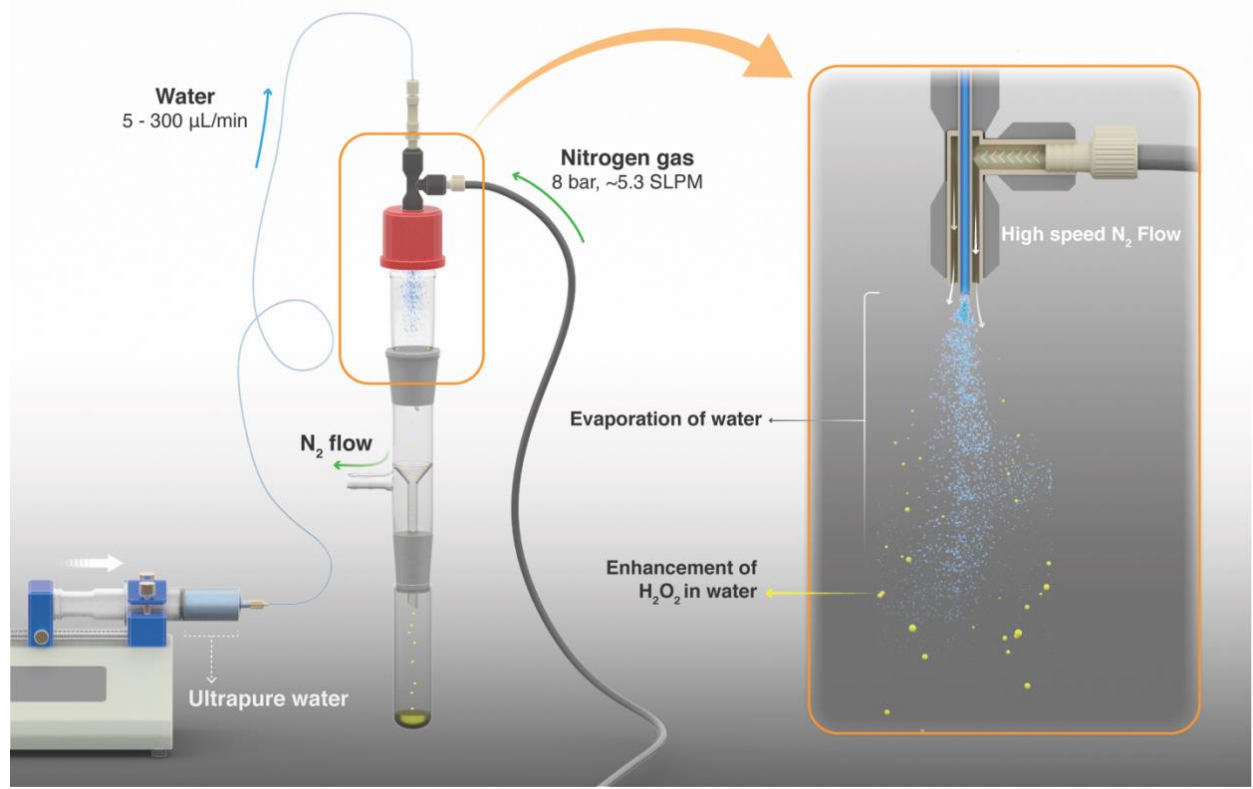

Fig. 1 – Schematic of the experimental setup illustrates the evaporative concentration of $H_2O_2$ in aqueous sprays. The spray setup is comprised of two coaxial stainless steel capillary tubes; liquid flows in the inner tube while nitrogen gas flows through the outer annulus (see 3D design and photographs in Fig. S2). Ultrapure water samples or aqueous solutions spiked with $H_2O_2$ are injected via a syringe pump and nitrogen gas is supplied by a high pressure cylinder. Sprayed microdroplets are collected in a clean glass flask until a sufficient analyte volume is collected (e.g., ≥400 µL analyte was needed for our HPAK analysis[29]). We define *Evaporation Ratio* as $V_O/V_F$, where, $V_0$ is the liquid volume injected volume to create the spray and $V_F$ is the volume of the collected sample. Note that the droplets on the walls as well as the analyted collecting at the bottom of the collector flask also contribute to *Evaporation Ratio*. Figure created by Heno Hwang, scientific illustrator at King Abdullah University of Science and Technology (KAUST).

**Evaporative concentration**

As a solution comprised of water and $H_2O_2$ evaporates, for instance, inside a vaccum oven or in the form of sprayed microdroplets, the ratio of the formed vapor ($H_2O(g):H_2O_2(g)$) is not the same as their bulk concentration ($H_2O(l):H_2O_2(l)$). This is so because water evaporates faster due to its lower boiling point (373 K at 1 atm) than that of $H_2O_2$ (423 K at 1 atm)[38, 39]. Here, we define *Evaporation Ratio* as the ratio of the initial volume prior to spraying ($V_o$) to the finally collected volume after spraying ($V_F$). A simple evaporation experiment was performed on bulk aqueous solutions containing 4 and 8 µM $H_2O_2$ by placing them inside a vacuum oven set at 20 mbar and 293 K for approximately 2 days. These solutions exhibited *Evaporation Ratios* of ~50 and ~17, respectively, such that the $H_2O_2$ concentration increased by ~12 and ~7-times, respectively. Next, we quantified *Evaporation Ratios* within pneumatic sprays formed via (i) ~0.5 µM $H_2O_2$ solutions and (ii) ultrapure water. The water flow rate was varied in the range of 50–400 µL/min and the nebulizing gas ($N_2$) flow rate was fixed at 5.3 L/min (~8 bar in-line pressure). Our experimental set up's details are presented in Figs. S2-S3 and Table S1; the influence of the in-line gas pressure and the inner/outer capillary dimensions on the gas flow rates is plotted in Fig. S3A-B; and the size distributions of sprayed microdroplets, determined via dynamic light scattering (Methods), as a function of gas flow is shown in Fig. S4B. Under these experimental conditions, when the liquid flow rate was > 70 µL/min, there was hardly any enhancement in the $H_2O_2$ concentration. However, as the liquid flow rate was reduced below 70 µL/min, the $H_2O_2$ concentration began to rise exponentially. Remarkably, within the range of 400–50 µL/min, the $H_2O_2$ concentration in the ~0.5 µM standard solutions increased by 9.1× (0.7 – 6.2 µM) and 13.6× (0.4 – 5.2 µM), respectively, and the *Evaporation Ratios* were 14.2 and 21.3, respectively. Similarly, the $H_2O_2$ concentrations in sprays formed with ultrapure water also increased when the liquid flow rate was reduced to the range 70–50 µL/min: the enhancement factors were 6.6× and 1.1× – both < 0.34 µM—while the *Evaporation Ratios* were 14.2 and 8.6, respectively. ANOVA tests confirmed that even the slopes of the ultrapure water samples (#1 and #2) were statistically greater than zero ($p < 0.05$). NB: the average time for sample collection (~1 ml for analysis) at 50 µL/min flow rate was > 5 h per datum point due to evaporation rates being almost as high as liquid injection rates; so, we did not pursue investigation of evaporation at lower liquid flow rates. Taken together, these experiments established that while evaporative concentration could enhance the $H_2O_2$ concentration in sprayed water microdroplets slightly, this effect is nearly a factor 100× too short to account for the ppm-level (1 ppm = 29.4 µM) $H_2O_2$ concentrations noted recently[26].

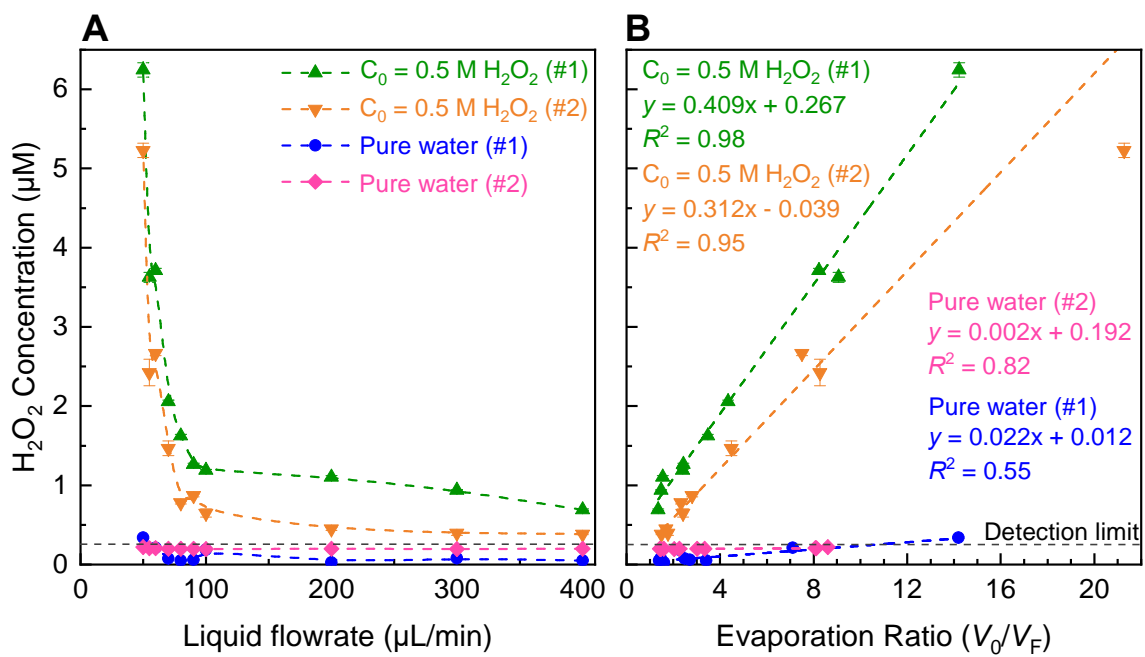

Fig. 2 – Evaporative concentration in pneumatic sprays formed with ultrapure water (blue and pink) and 0.5 μM $H_2O_2$ solutions (green and orange) using a fixed $N_2$ gas flowrate 5.3 L/min. (A) Sharp rise in the $H_2O_2$ concentration of (spiked) 0.5 μM solution when the liquid flow rate was < 70μL/min; minimal $H_2O_2$ production in sprays formed with ultrapure water. (B) Relationships between the final $H_2O_2$ concentration and the evaporation ratios. Note: the error bars correspond to the standard deviations in the spectrophotometer readings. All slopes in (B) were statistically greater than zero ($p < 0.05$; ANOVA test).

**Mechanical vibrations and shock waves**

We combined experiment and theory to probe the effects of mechanical vibrations on the formation of shock waves during spraying. High-speed videos of pneumaticly driven sprays were recorded using a Kirana-05M camera at 5 million frames per second (fps). Its 200 ns time-resolution enabled us to observe microdroplets trajectories before, during, and after their interaction with the shearing gas (Fig. 3A and SI Movie S1); their size, number, and velocity were also monitored (Methods). We found that as the slow-moving water droplets got hit by the fast-moving gas, they accelerated and frequently broke into smaller ones (Fig. 3A). We deduced that the speed of the $N_2$ gas to be ~800 ms$^{-1}$.

Before computational fluid dynamic (CFD) simulations, a theoretical calculation was conducted based on the Rankine-Hugoniot conditions, because from a CFD simulation point of view $H_2O_2$ could only be generated by high-temperature reactions (Section S1). Shock waves could generate high temperatures and pressures. When the high-speed gas hits the water droplet, almost all of the momentum energy is converted to heat. If we assume the heat capacity does not change, then, the temperature rise of the static gas can be estimated as, $\frac{u^2}{2}/c_p$, where $u$~800 ms$^{-1}$ and $c_p$= 4.2 J/g-K. This yields a ≈301 K rise in temperature at the droplet front.

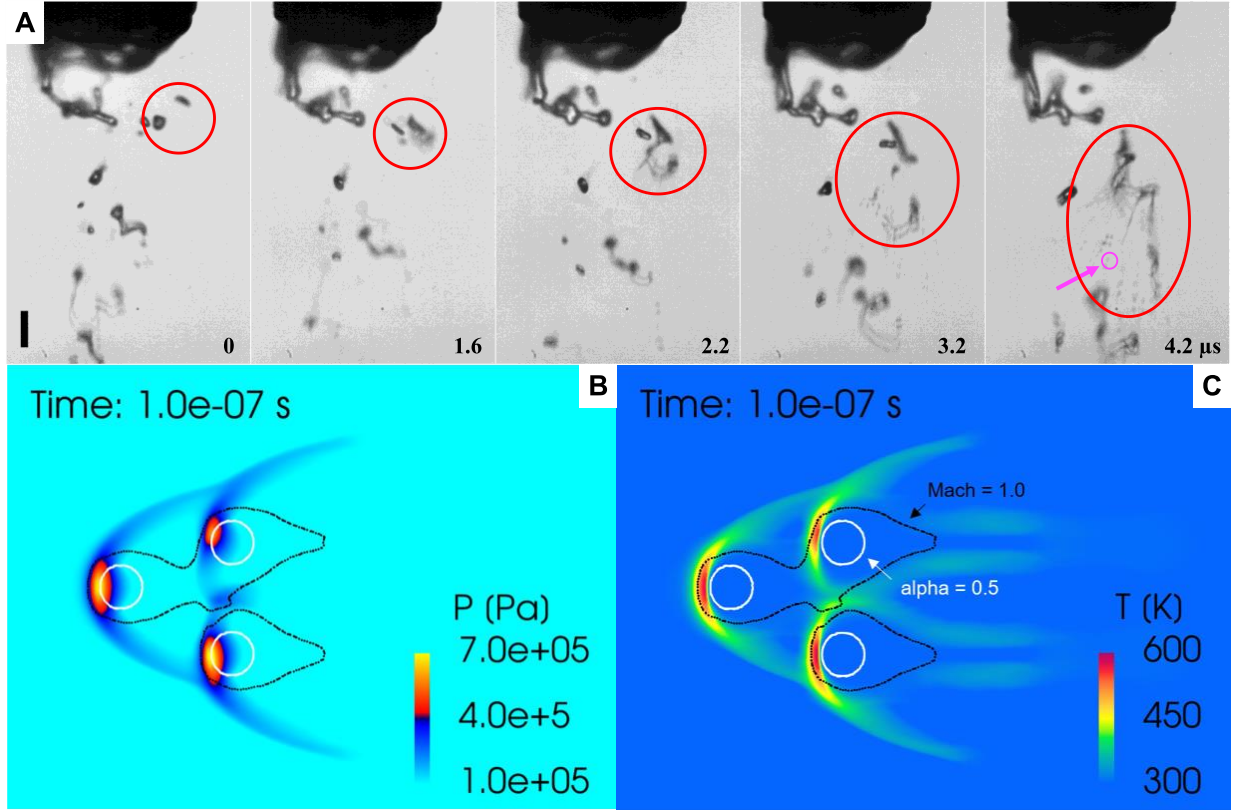

Fig. 3 – (A) A representative sequence of video images showing interactions between concentric high-speed external nitrogen gas (5.3 L/min) and a slower inner water jet (100 µL/min). Red circles indicate a collision and breakup of two droplets upon acceleration by the fast-moving nitrogen gas. Pink arrow shows a droplet of 4 µm diameter. The images are acquired with an ultra-fast camera (Kirana-05M) at 5 million fps using a Leica long-distance microscope. The sequence of frames are shown at times relative to the first frame, t = 0, 1.6, 2.2, 3.2, and 4.2 µs. The scale bar in panel (A) is 50 µm. (B-C) Predicted distributions of pressure and temperature fields from the numerical simulations of high-speed gas-flow around three adjacent droplets of 20 µm diameter.

Next, we probed the gas-water interactions via three-dimensional CFD simulations using the Converge code, wherein the turbulence is simulated by the renormalization group k-ε model[45]. The Eulerian void of fluid (VOF) method [46] was adopted to capture the in- and near-nozzle spray details (SI Section S1 and Figs. S5–S6). Simulations revealed the formation of bowl-shaped shock waves at the impact point of the droplets that led to an increase in the local temperature (~327 °C) and pressure (~7 bar) (Figs. 3B–C). However, these conditions are too mild to produce a chemical transformation; a relatively high concentration of $H_2O_2$ was generated only with a temperature over 1000 K and a residence time over 10 μs (Fig. S7) that are unachieveable in our experiments. Thus, this hypothesis cannot explain the formation of $H_2O_2$ in sprayed water microdroplets.

**Effects of Atmospheric Ozone on the $H_2O_2$ Formation in Water**

Next, we exploited a glovebox to investigate the effects of ozone ($O_3$) gas concentration in the air on the $H_2O_2$ formation in water microdroplets (Fig. 4). The $O_3$ concentration in the air was varied in the range 2–4900 ppb and measured using an ozonometer with a 2 ppb detection limit and a range of 2–5000 ppb (Methods). Subsequently, the $H_2O_2$ concentration in the water microdroplets was quantified using the HPAK.

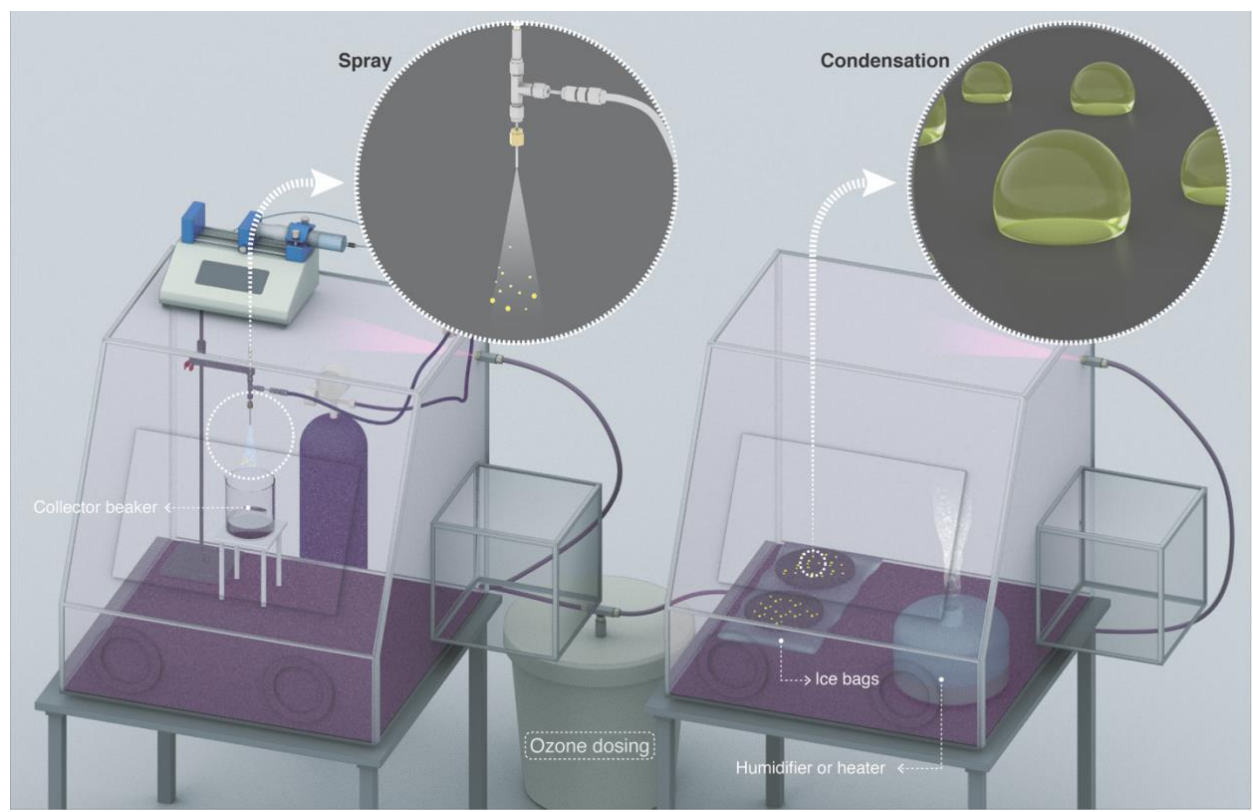

Fig. 4 – Schematics of experimental set-up for conducting spray and condensation experiments inside glovebox with controlled gas composition. Ozone gas is produced outside and diluted with $N_2$ gas and air before its entry inside the glove box, where its concentration is monitored using an ozonometer with a 2 ppbv detection limit. On the left-side, we illustrate the spray formation and on the right-side, we illustrate water vapor condenstation on hydrophobic substrates. Figure created by Heno Hwang, scientific illustrator at KAUST.

Below, we discuss the various scenarios in this chronology: (i) condensing vapor formed using heated water under ~85% RH (Methods) (ii) condensing water from a commercial ultrasonic humidifier with 15 W output power and RH varying in the range 70-80% (Methods), and (iii) pneumatic sprays formed with a water flow rate of 1 mL/min and shearing gas ($N_2$) flow rate of 2.3 L/min. Single crystal $SiO_2$/Si wafers silanized with perfluorodecyltrichlorosilane (FDTS) using a molecular vapor deposition technique[47] were used as substrates. For the condensation experiments, the substrates were cooled down to 3–4 °C. Next, it is crucial to note that the $O_3(g)$ concentration in our laboratory as well as at KAUST campus during the course of this study was < 2 ppb. Thus, the glovebox filled with ambient air served as the control case. In this scenario, the water collected from the pneumatic sprays or the condensates formed via the vapor of hot water (50–70 °C) did not contain $H_2O_2$ (Fig. 2 – blue and pink data points and Ref.[29]). Curiously, when the water vapor was derived from a 15 W commercial ultrasonic humidifier, ~1 µM $H_2O_2$ was detected in the condensate. We have explained its origins in a recent report[29] as well as in the introduction above.

$O_3(g)$ was produced using an ozone generator and diluted with air (or $N_2$ gas) prior to its entry in the glovebox equipped with an ozonometer (Methods). In the condensation experiments, as the $O_3(g)$ and the RH reached the desired levels, the $O_3(g)$ supply was terminated (SI Section S2). We monitored the time-dependent loss of $O_3(g)$ concentration as we waited for 40 minutes to collect adequate condensate volume to be able to perform the HPAK analysis. In some cases the $O_3(g)$ concentration fell below the detection limit (Fig. S8). Fig. 5 (green and red datapoints) presents the final $H_2O_2$ concentration in the condensates (from the humidifier and hot water, 40 °C) against the initial $O_3(g)$ concentration. NB: these data should not be mistaken to represent thermodynamic equilibrium, which is not achieved in our system. Specifically, a number of factors influence $O_3(g)$ depletion, e.g., reactions with and/or adsorbion onto the glovebox surface and materials/instruments inside it, reactions with condensed water on the substrates as well as the cold surfaces underneath them, and the leakage from the glovebox. To appreciate this further, we point out that when a metallic heating plate was used to produce water vapor inside the glovebox for condensation experiments, $O_3(g)$ depleted rapidly (Fig. S8B). In comparison, when an ultrasonic humidifier with a plastic surface was utilized, the decay was gradual. This also is the reason for the lower $H_2O_2$ (aq) concentration in the condensates of heated water in comparison with that of the humidifier (Fig. 5).

To distinguish the $O_3$ consumption by the water from the parasitic losses in the glovebox, we compared the time-dependent depletion of $O_3(g)$ with and without water. In this experiment, 120 mL of water was placed inside a shallow containment presenting 390 cm$^2$ of the air–water surface area. The $O_3(g)$ depletion was faster in the presence of water (Fig. S9). These experiments and the controls establish unambiguously that (i) water contributes to the $O_3(g)$ depletion and (ii) the higher the initial $O_3(g)$ concentration, the higher is the formation of $H_2O_2$ in the condensates (Fig. 5 – green and red datapoints). We explain in the Discussion section what happens to the depleted $O_3(g)$ inside water.

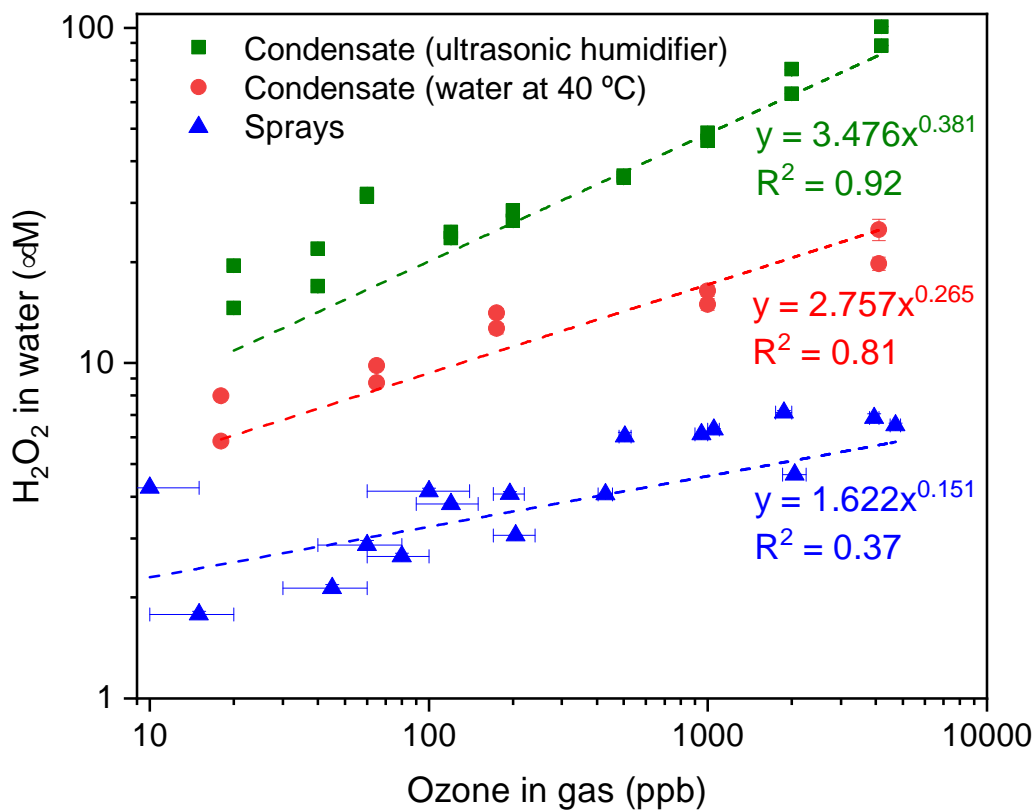

**Fig. 5** – $H_2O_2$ concentration in water microdroplets formed inside the glovebox via condensation from two humidity sources, (i) ultrasonic humidifier (18-20 °C; RH 70-80%) and (ii) water heated beaker (40 °C; RH ~85%), and via pneumatic sprays of pure water at 100 µL/min with 2.3 L/min $N_2$ flow rate (18-20 °C; RH 30-60%). As the $O_3$ concentration in the glovebox atmosphere increased, the $H_2O_2$ concentration in the microdroplets also increased. The condensation experiments were performed by setting the initial concentration in the gas phase and then stopping all in and out flows, this led to the gradual depletion of ozone from the glovebox; while for the spray was performed by controlling the ozone concentration with manual adjustments in the dilution of ozone by nitrogen gas and clean air (Fig S8). Horizontal error bars in the sprays represent the range of ozone concentration fluctuation during the time of sample collection.

Next, we discuss our spray experiments under controlled $O_3(g)$ environment (Figs. 4(left) and S2). To offset the dilution of the gas-phase $O_3(g)$ due to the shearing gas flow from the sprays ($N_2$, 2.3 L/min at NTP), we manually controlled the flow of $O_3(g)$ and air, now, via separate lines (SI Section S2). Thus, we managed to achieve relatively stable $O_3(g)$ concentration (Fig. S8C) and RH in the range of 30-60%. At the above-listed water and shearing gas flow rates, we were able to collect adequate sample volumes for the HPAK analysis within 5 min. Fig. 5 (blue datapoints) present the $H_2O_2$ concentration in the water collected from the sprays; note- in these experiments, the $O_3(g)$ does not deplete with time (Fig. S8C) unlike the case for the condensation experiments (Fig. S8A-B). To probe the effects of the sprayed microdroplet size distribution on the formation of $H_2O_2$, we varied the shearing gas flowrate while keeping the water flow rate at 1ml-min$^{-1}$ and the $O_3(g)$ concentration fixed at 100±20 ppb. As the shearing gas flow was varied from 1–5 L-min$^{-1}$, the mean droplet diameter decreased non-linearly within the range 105–14 μm (Fig. S4A–B). However, the the $H_2O_2$ concentration increased linearly with nitrogen flowrate (Fig. 6). This indicates that, despite the higher surface, the main reason for the increase in $H_2O_2$ concentration with nitrogen flowrate has to do with enhanced mixing between the low-ozone spray region and with the surrounding ozone-containing atmosphere in the glovebox. Some other factors that influence the droplet-$O_3(g)$ interactions and the $H_2O_2(aq)$ formation include, the life-span of the sprayed water droplets, the $O_3(g)$ concentration at the droplet surface and its localized depletion/repletion, the droplet surface-area-to-volume ratio, $O_3$ influx into the water, and the $O_3$ and $H_2O_2$ outflux from the microdroplets. It should also be realized that the blue datapoints in Figs. 5–6 do not ascribe to the liquid–gas thermodynamic equilibrium, which is a common characteristic of sprays.

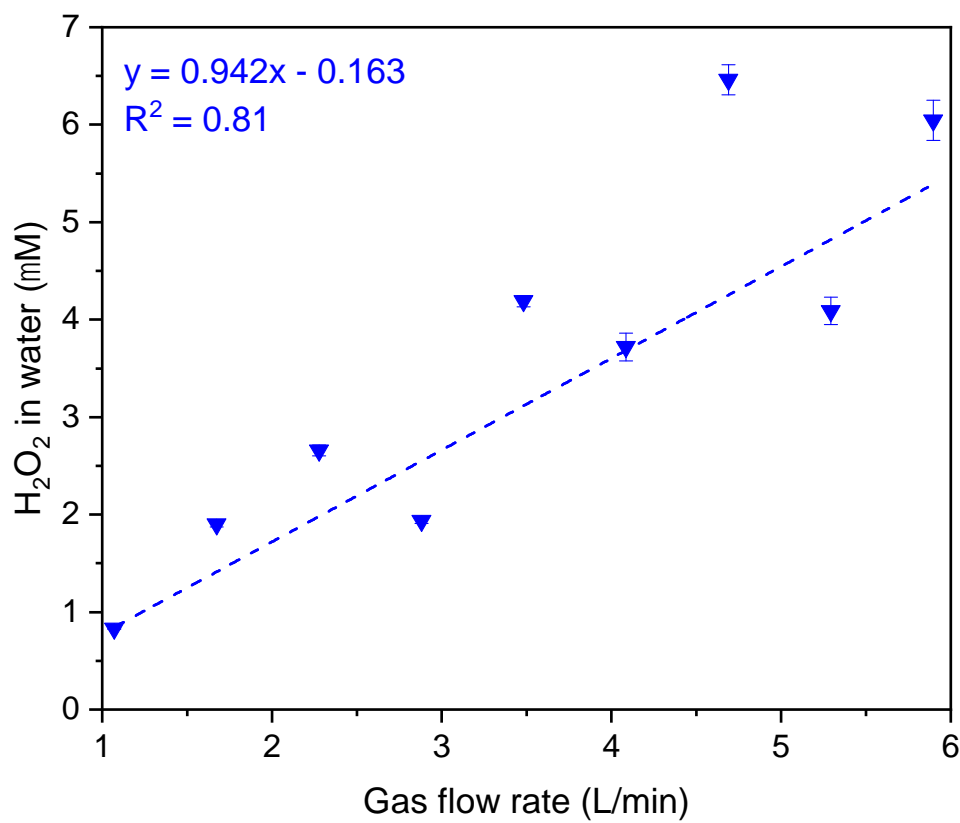

Fig. 6 – Influence of nitrogen gas flow rates on the formation of $H_2O_2$ in water microdroplets produced by pneumatic sprays under ozone atmosphere (100±20 ppbv). The liquid flow rate was fixed at 1 ml/min.

**Discussion**

Here, we draw together our results to pinpoint the key factors and mechanisms responsible for the $H_2O_2$ formation in water microdroplets. First of all, the fact that we do not see the $H_2O_2$ formation in water microdroplets, sprayed or condensed, in the absence of $O_3$ proves that this chemical transformation is not a property of the air–water surface (Fig. 2 – blue and pink data points and Ref.[29]). While the evaporative concentration of trace quantities of $H_2O_2$ already present in water is possible, in principle, in sprays, neither is it adequate to explain the ppm-level $H_2O_2$ concentration in the previous report[26] nor is it relevant to the condensation experiments[27]. Similarly, shock waves produced during pneumatic spraying under our experimental conditions simply cannot contribute to ppm-level $H_2O_2$; also, they are irrelevant to condensation-based reports. These experiments, however, facilitate valuable insights into relevant physical and chemical processes that could, under extreme conditions, contribute to the $H_2O_2$ concentration/formation in water.

Next, we discuss the scope and significance of our investigation of the $H_2O_2$ formation in condensed and sprayed microdroplets under controlled $O_3(g)$ environment. These experiments unambiguously pinpoint that the $H_2O_2$ formation in water is extremely sensitive to the ambient $O_3$ concentration (Figs. 5 and S8); if the $O_3(g)$ is present, then the enhancement in the air–water surface area combined with higher mixing with the surrounding ozone containing atmosphere leads to higher $H_2O_2$ formation (Figs. 6 and S4). Based on this insight, below, we present an alternative explanation for the $H_2O_2$ formation phenomenon reported by researchers based in California, i.e., the original reports[26, 27]. The Environmental Protection Agency (EPA)[48] live database reports that the average $O_3$ concentration in California for the year 2020 was ~32 ppbv with a maximum daily average of ~43 ppbv; the highest $O_3(g)$ concentrations on some days could even exceed 80 ppb. <u>We submit that this high regional $O_3(g)$ in California is the real reason behind the $H_2O_2$ formation in the original reports[26, 27]</u>. Next, we provide some semi-quantitative insights into this chemical transformation.

It is well-known that $O_3(g)$ is minimally soluble in water (Henry's law constant, $H_{O_3}^{CP} = 10^{-2}$ Mol/L-atm at NTP)[44]. Therefore, in our experiments, the aqueous ozone concentration would always be lower than the equilibrium concentration specified by the Henry's law, i.e., $C_o = P_{O_3} \times H_{O_3}^{CP} \approx 43 \times 10^{-9} \times 10^{-2} = 0.4$ nM. However, the following factors can lead to the reported 30–115 μM $H_2O_2$ concentrations:

1. Microdroplets enhance the surface-to-volume ratio dramatically. E.g., if a spherical water droplet of volume 1 mL and diameter $D = 1.24$ cm is sprayed into microdroplets of diameter $d = 10$ μm, the surface area increases by a factor of $D/d = 1240$. This enhancement in the air–water surface area facilitates significant $O_3(g)$ influx into the water leading to $O_3(aq)$ despite the lower non-equilibrium concentration. Additionally, discrepancies between the concentrations in sprays reported here versus those in the original reports[26, 27] are attributed to differences in spray setups dimensions. This leads to different gas flowrates (Fig. S3) and aerodynamic conditions that influence the mass transfer of ozone to the liquid and the formation of $H_2O_2$(Fig. 6).

2. $O_3$(aq) undergoes fast autodissociation reactions in water leading to a variety reactive species such as OH, $HO_2$, $HO_2^-$, $O^-$, $O_2^-$, OH, and $O_3^-$ and a stable product, i.e., $H_2O_2$[43]. For instance, the following reactions could explain this mystery:

$$O_3(g) \rightarrow O_3(aq) \qquad (Eq.1)$$
$$O_3(aq) + OH^-(aq) \rightarrow HO_2^-(aq) + O_2(aq) \qquad (Eq.2)$$
$$HO_2^-(aq) + H_3O^+(aq) \rightarrow H_2O_2(aq) + H_2O(aq) \qquad (Eq.3)$$

NB: other reaction pathways could also contribute to the $H_2O_2$ formation. Now, in contrast to ozone, $H_2O_2$ is miscible in water; so, a significant fraction of the $H_2O_2$ formed in the water should stay within ($H_{H_2O_2}^{CP} \sim 10^4$ M/atm at NTP). In the presence of organics in water, such as in the natural and industrial contexts, products of ozone autodissociation participate in numerous reactions and get consumed[49, 50]; solid surfaces such as silica and perfluorocarbons can also influence the rates of these reactions[43]; we also note that ozone generators could produce NOx species that could interfere with reactions[51]. However, an in-depth analysis of these factors as well as the rates of reactions and mechanisms, the mass transfer considerations, and the non-equilibrium effects listed in the results section fall beyond the scope of this study. We close the discussion by noting that this $O_3$–$H_2O_2$ coupling inside water has led to misinterpretations in the past also[43, 49, 52] and strategies for curtailing these effects have been devised[36]. Furthermore, for a reader interested in simulations of the mass transfer of $O_3$(g) and $H_2O_2$(g) at the water surface and subsequent chemistries, we refer to this excellent report[53].

**Conclusion**

In this contribution, we investigated the factors and mechanisms responsible for the formation of $H_2O_2$ in water microdroplets produced via spraying or condensation. We found that in the absence of ozone gas, $H_2O_2$ does not form regardless of condensation/spraying or the droplet size or the substrate. This unambiguously establishes that the air–water interface does not spontaneously produce $H_2O_2$. The $H_2O_2$ formation in microdroplets only happens in the presence of $O_3$(g); as the $O_3$(g) concentration increases in the air, the $H_2O_2$ concertation also increases in the microdroplets – condensed or sprayed. The higher the gas–liquid surface area, mixing, and contact time, the higher is the $O_3$(g) uptake in the water; this $O_3$(aq) then undergoes chemical transformations yielding $H_2O_2$(aq). Through the Environmental Protection Agency (EPA) database[48], we found that the average $O_3$ concentration in California for the year 2020 was ~32 ppbv, with a maximum daily average of ~43 ppbv. These facts and our findings therefore resolve the mystery of the spontaneous formation of $H_2O_2$ in water microdroplets reported from California. The primary role of the air–water interface is to facilitate the mass transfer of ozone into the water. Of course, reactions of $O_3$ with the water surface could be implicated, and they should be probed via surface-specific platforms in conjunction with theory and computation. To sum up, our findings disprove the latest claims for the water surface's ability to spontaneously produce $H_2O_2$ [26, 27]. The speculation for the presence of the mysteriously high electric fields at the air–water interface, responsible for transforming $H_2O$ into $H_2O_2$, thus also appears untenable. Lastly, our findings can also help explain the origins of the recently reported bactericidal properties of pneumatically sprayed water microdroplets and help assess their practical and environmental relevance[54, 55]. Since $H_2O_2(aq)$ can participate in both oxidation and reduction reactions[56], we close by proposing a re-examination of recent reports on unexpected reduction[57-59] and oxidation[60] reactions realized in aqueous sprays.

## Materials and Methods

### Chemicals

Standard hydrogen peroxide ($H_2O_2$) 30% and high-performance liquid chromatography (HPLC LC-MS) grade water were purchased from VWR Chemicals (Catalogues #270733 and #23622.298). Deionized water produced from a MilliQ Advantage 10 set-up was also used in this study.

### Quantification of $H_2O_2$ in water
### Hydrogen Peroxide Assay Kit (HPAK) assay

$H_2O_2$ concentration present in both condensed and sprayed water microdroplets was quantified by the Hydrogen Peroxide Assay Kit (Fluorometric-Near Infrared, Catalogue #ab138886). It can detect $H_2O_2$ by the fluorescence produced when in contact with the AbIR Peroxidase Indicator, and its maximum excitation and emission wavelengths are 647 nm and 674 nm, respectively. This method also contains a Horseradish peroxidase enzyme that catalyzes the reaction and increases the fluorescence signal, having a linear range of detection from 250 nM to 10 µM. Samples containing higher concentrations than the detectable range were diluted with deionized water. The analyses were conducted in a 96-well black/clear bottom microtiter-plate, adding 50 µL of each sample within 50 µL of HPAK, using a SpectraMax M3 microplate reader (Molecular Devices LLC) and the software SoftMax Pro 7 for fluoresce reading. The $H_2O_2$ concentration was calculated using the calibration curve obtained on the same day.

### Peroxide test strips for semi-quantitative analysis:

We used peroxide test strips (Baker Test Strips, VWR International) with a detection limit of 1 ppm (29.4 µM) for a qualitative estimation of $H_2O_2$ in aqueous samples. They work through a colorimetric reagent which gives blue color when exposed to $H_2O_2$(aq) and the color deepens with the $H_2O_2$ concentration.

### Water Microdroplets Generation via Sprays

We built a spray to generate water microdroplets by injecting water through an inner tube of 100 µm of diameter using a syringe pump (PHD Ultra, Harvard Apparatus), and nitrogen coaxial sheath gas through a 430 µm diameter tube, both tubes made of stainless steel. For the concentration by evaporation experiments, HPLC grade water was used, and a glass reactor (a tube of three glass pieces connected with only a small opening for gas release) was connected to the spray to reduce the evaporation while collecting all the sprayed droplets. The water flow rate used varied from 50 to 400 µL/min , while the nitrogen gas flow rate was set at 5.3 L/min. For the experiments inside the glovebox, deionized water was injected at a flow rate of 1 mL/min and the nitrogen gas flow rate varied from 1.1 to 5.9 L/min, with most of the experiments set at 2.3 L/min.

**Sprayed Water Microdroplets Diameter Acquisition**

The distribution of microdroplets-size was measured with a Spraytec system (Malvern Instruments). The interaction between the laser beam with the spray produces a diffraction pattern, from which the derived parameter Sauter Mean Diameter (SMD) was calculated. Our spray was positioned at ~2 cm from the laser.

**Substrates for Condensation**

Silicon wafers (p-doped, <100> orientation, 4" diameter, thickness of 500 µm and a 2 µm-thick oxide layer) were purchased from Silicon Valley Microelectronics (Catalogue #SV010).

**Functionalization of $SiO_2$/Si wafers**

To make our $SiO_2$/Si wafers hydrophobic, we functionalized them by silanization with perfluorodecyltrichlorosilane (FDTS). Firstly, we removed any organic contaminants and hydroxylated the surface by trating it with oxygen plasma for 2 min. Then by using a molecular vapor deposition process (Applied Microstructures MVD100E), we grafted our silicon wafers with FDTS by applying one of our previous methodologies reported[61].

**Ozone Generation and Experiments inside the Glovebox**

A portable isolated glovebox (Cleatech, Catalogue #2200-2-B) was used as a controlled-environment chamber for ozone concentration. An ozone generator (Mainstayae; $O_3$ production rate: 24 g/h) was placed outside the glovebox inside a bucket with a hole for plastic tubing with an air flow. To control the ozone concentration inside the glove box, this mixture of air and ozone was further mixed with a different channel containing more air and nitrogen, all of them controlled by valves to change the gas flows. A portable ozone meter (GoolRC) with detection range of 2–5000 ppbv and detection limit at 2 ppbv was positioned inside the glovebox for real time monitorament; its proper functioning required RH < 85%. When the ozone concentration was reached, varying from 10 to 4900 ppbv, the silicon wafers used as substrates for water microdroplets formation were placed onto an ice–water bag with uniformily distributed temperature at ~3 °C, which quickly achieved thermal equilibrium, and left for 40 min of condensation time. The temperature on the substrates was checked via a non-contact digital infrared thermometer (Lasergrip 774). In the case of the sprays, the nitrogen gas was already flowing considering its great effect on the ozone concentration inside the glove box, and as soon as the ozone concentration was stabilized, the water injection was started for collection of the microdroplets into a glass funnel inside a beaker during 3-13 min. The relative humidity inside the chamber was kept in the range 70–80% using the ultrasonic humidifier, at ~85% by heating water and 30-70% when using the spray (a bigger range as the air and nitrogen gas flows were affecting it more). The ambient air temperature was in the 18–20 °C range. To collect the samples, we poured the droplets from the silicon wafers into a clean glassware, and all the samples were transfered into a 15 mL centrifuge tube (VWR International).

**Water Vapor Generation via Ultrasonic Humidifier**

An ultrasonic humidifier (Proton PHC 9UH) with 15W of power was used in this study. This equipment produces mist from water by ultrasonic waves generated from the piezoelectric disk located on its bottom. Deionized water was used and the ultrasonic humidifier was positioned far away from the silicon wafers (~40 cm apart) to avoid direct mist deposition.

**Water Vapor Generation via a Heating Plate**

Deionized water was heated at 40 °C using a heating plate (IKA RCT, Catalogue #3810000). The plate was positioned ~30 cm apart from the substrates, and the temperature was controlled by the coupled temperature sensor (PT 1000.60), which was in contact with the water.

**High Speed Imaging**

The rapid interaction between nitrogen gas and water requires observation with an ultra-high-speed video camera (Kirana-05M, Specialized Imaging, Tring UK) acquiring 180 images at capture speeds of up to 5×10$^6$ fps, with a full resolution of 924(W) x 768(H) px irrespective of the frame rate used and magnification with Leica long-distance microscope at magnification up to 29.4. The short- and long-term dynamics of this interaction can be captured using several different frame rates. The framing is synchronized with red diode-lasers with adjustable pulse duration of 50–170 ns per frame to minimize motion smearing. The smallest detectable diameter of the droplets is around 4 µm.

**Computational Methods**

Three-dimensional (3D) CFD simulations were performed using the Converge code to simulate the sprays. The turbulence was simulated by the renormalization group k-ε model[45]. The Eulerian void of fluid (VOF) method [46] was adopted to capture the in- and near-nozzle spray details. In this method, the gas and liquid fuel are considered as a single compressible fluid mixture, and the void fraction ($\alpha_l$) is used to represent the volume fraction of liquid. Details of the related models are available in the reference[62].

# Supplementary Information

## On the Formation of Hydrogen Peroxide in Water Microdroplets

Adair Gallo Jr.[1,&], Nayara H. Musskopf[1,&], Xinlei Liu[2], Ziqiang Yang[2], Jeferson Petry[1], Peng Zhang[1], Sigurdur Thoroddsen[2], Hong Im[2], Himanshu Mishra[1,*]

[1]Interfacial Lab (iLab), Biological and Environmental Science and Engineering (BESE) Division, Water Desalination and Reuse Center (WDRC), King Abdullah University of Science and Technology (KAUST), Physical Science and Engineering (PSE), Thuwal 23955-6900, Saudi Arabia

[2]Physical Sciences and Engineering (PSE) Division, King Abdullah University of Science and Technology (KAUST), Physical Science and Engineering (PSE), Thuwal 23955-6900, Saudi Arabia

[&]Equal author contribution

*Himanshu.Mishra@kaust.edu.sa

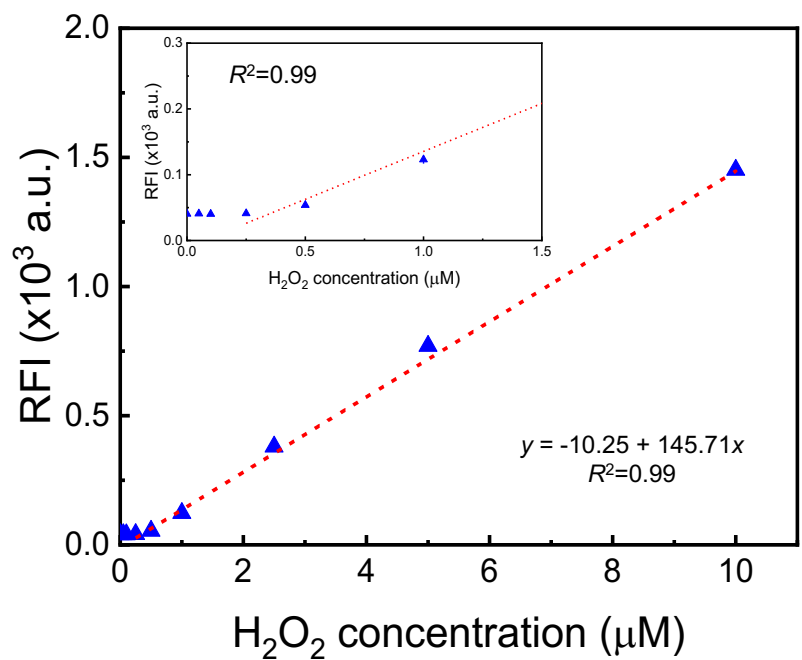

Fig. S1 – Calibration curve for $H_2O_2$ concentration using the Hydrogen Peroxide Assay Kit (HPAK) with fluorescence maximum absorption and emission at 647 and 674 nm, respectively. It is identical with our previous report[1].

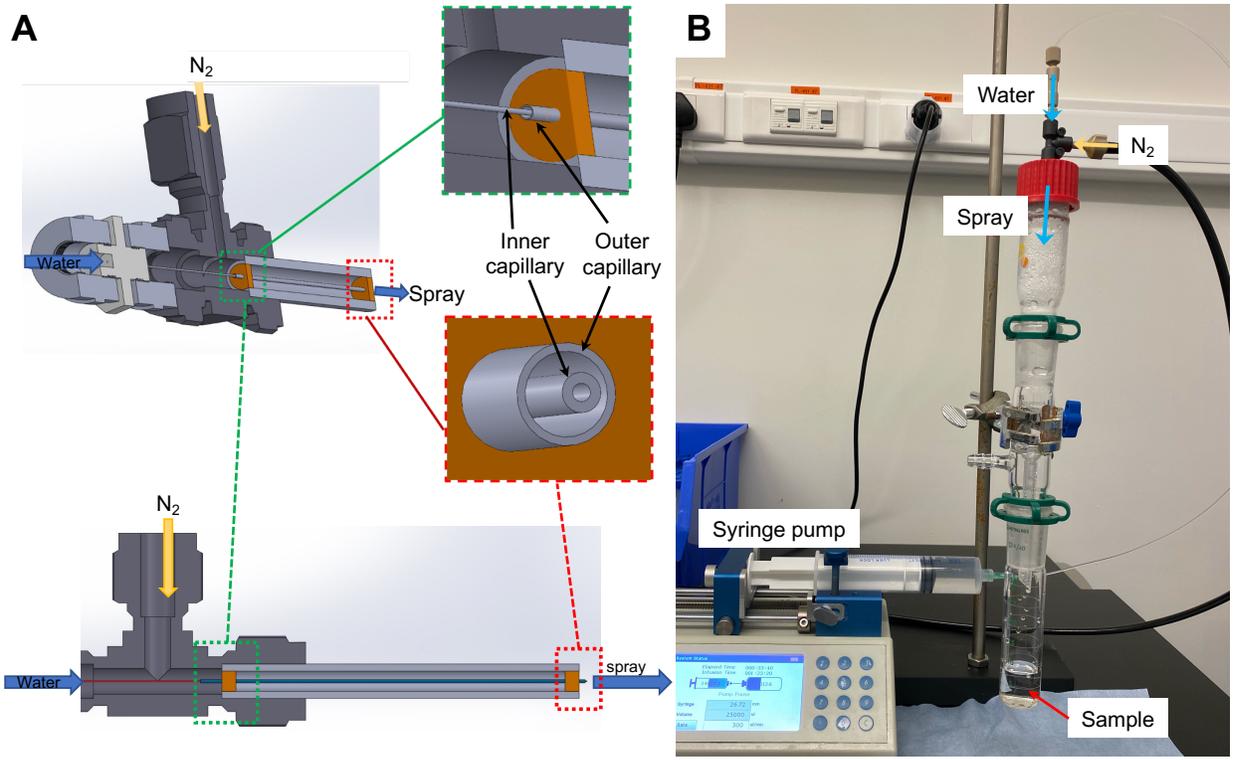

Fig. S2 – (A) Schematics of pneumatic spray setup. (B) Photograph of spray setup connected to a glass flask for sample collection.

Table S1 – Dimensions of a few models of our custom-built pneumatic spray setups

| Spray type | Inner capillary (±5 µm) | | Outer capillary (±5 µm) | | Coaxial | $N_2$ flow cross-section |
|---|---|---|---|---|---|---|
| | Inner dia. (µm) | Outer dia. (µm) | Inner dia. (µm) | Outer dia. (µm) | Length (±1 mm) | Area (µm²) |
| **Spray A (main)** | 100 | 228 | 432 | 585 | 24 | 421663 |
| Spray B | 106 | 222 | 489 | 587 | 24 | 596554 |
| Spray C | 113 | 217 | 404 | 712 | 46 | 366282 |
| Spray D | 89 | 213 | 473 | 731 | 79 | 562181 |

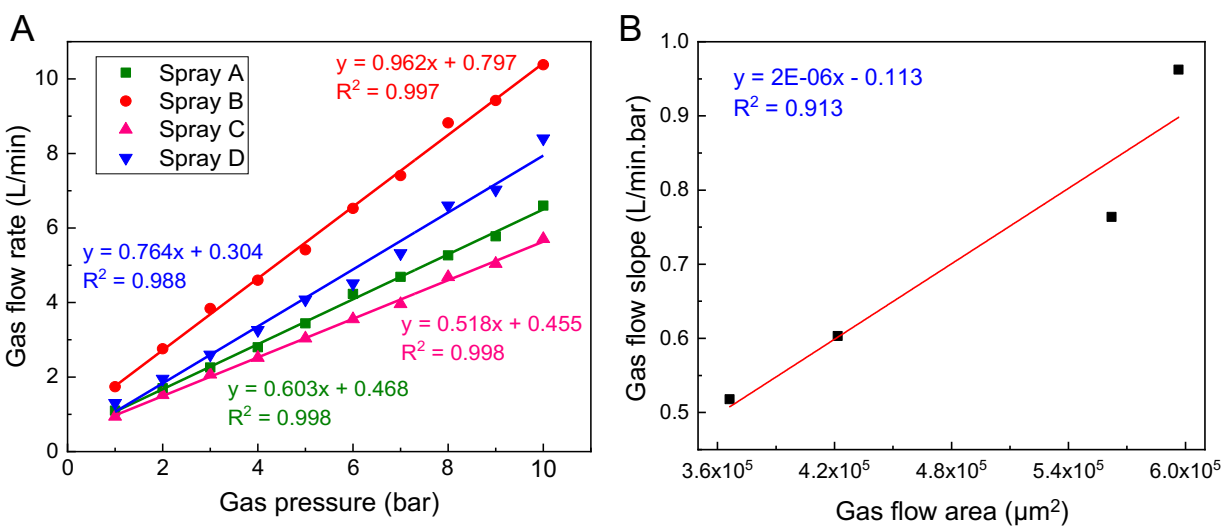

Fig. S3 – (A) Dependence of gas flow rate in-line gas pressure for four different types of custom-built sprays, whose dimensions are shown in Table S1. (B) Linear correlation between the slopes of gas flow rate by in-line gas pressure (from A) and gas flow area of the four different sprays.

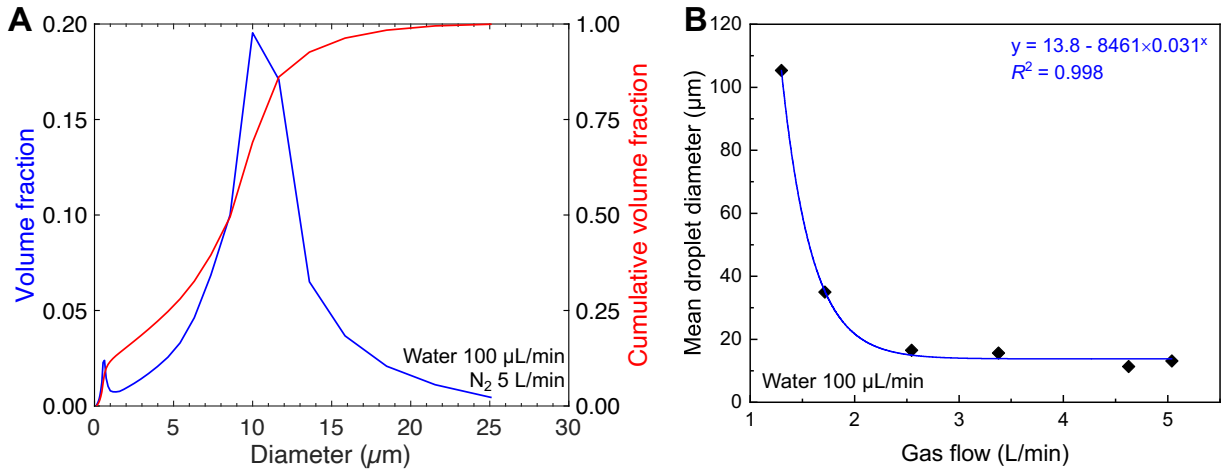

Fig. S4 – (A) Droplet size distribution and (B) droplet mean diameter as function of gas flowrate in sprays.

## Section S1. Theoretical and computational section

Before computational fluid dynamic (CFD) simulations, a theoretical calculation was conducted based on the Rankine-Hugoniot conditions, because from a CFD simulation point of view $H_2O_2$ could only be generated by high-temperature reactions. To generate high temperatures, the shock wave is a possible approach. By using a high-speed $N_2$ gas, the interaction between $N_2$ and a relatively static water jet could lead to shock waves. In the experiment, a high-pressure (8.27 bar) co-flow $N_2$ gas was employed. By using the momentum and energy conservation equations,

$$\rho_1 u_1^2 + p_1 = \rho_2 u_2^2 + p_2 \quad (1)$$
$$h_1 + u_1^2/2 = h_2 + u_2^2/2 \quad (2)$$

We know that the exit velocity of the high-speed gas is about 792 m/s at room temperature. When the high-speed gas hits the water droplet, almost all of the momentum energy is converted to heat. If we assume the heat capacity does not change. Then the temperature rise of the static gas is $\frac{u_2^2}{2}/c_p$, which is about 301 K. Schematic of the process is depicted in Fig. S5.

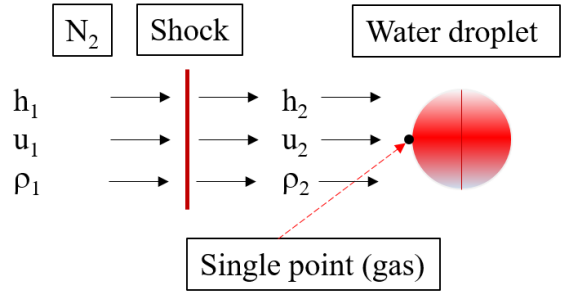

Fig. S5. Schematic of the interaction between high-pressure $N_2$ gas and static water droplet.

To further clarify the gas-water interaction process, three-dimensional (3D) CFD simulations were performed using the Converge code. The turbulence is simulated by the renormalization group k-ε model [2]. The Eulerian void of fluid (VOF) method [3] was adopted to capture the in- and near-nozzle spray details. In this method, the gas and liquid fuel are considered as a single compressible fluid mixture, and the void fraction ($\alpha_l$) is used to represent the volume fraction of liquid. Details of the related models are available in [4]. Fig. S6 illustrates the computational domain (200 μm in diameter and 400 μm in length). To mimic the experiment, the inlet boundary was imposed with a high-speed $N_2$ gas was imposed and three droplets with a diameter of 20 μm were scattered in the central domain. A base mesh size of 5 μm was used and a fixed embedding region was adopted with a refined scale of 4, which yields a minimum mesh size of 0.625 μm. A varying time step was used by controlling the convective Courant flow number to be below 0.5.

Fig. 3 shows the predicted distributions of pressure and temperature flow fields at 0.1 μs. Note that the pressure jumps to a significantly high level when the high-speed gas impinges onto the static water droplet, which results in a high-temperature rise of about 300 K, in agreement with the theoretic calculation.

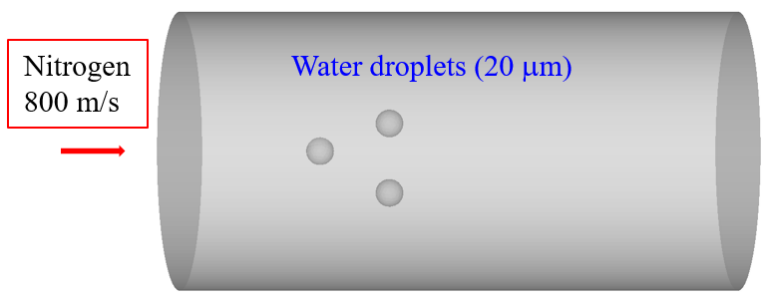

Fig. S6. Schematic of the computational domain.

However, a 300 K temperature rise might not be enough to explain the significant growth of $H_2O_2$ production during the experiment. Further zero-dimensional simulations were conducted to evaluate the effects of temperature on $H_2O_2$ [5] formation using the SENKIN code. In simulations, a homogeneous constant volume reactor was employed with pure gas water as the single reactant. The boundary was at constant pressure (1 atm) and temperature conditions. Various temperatures and mixture residence times were studied and the final $H_2O_2$ production was extracted for further analyses. Fig. S7 shows the predicted $H_2O_2$ concentration at various temperatures and residence times. Expectedly, almost no $H_2O_2$ was generated at low temperatures and residence times; a relatively high concentration of $H_2O_2$ was generated only with a temperature over 1000 K and a residence time over 10 μs. However, it should be noted that the concentration is only at an order of $10^{-11}$ even at a high temperature and a long residence time.

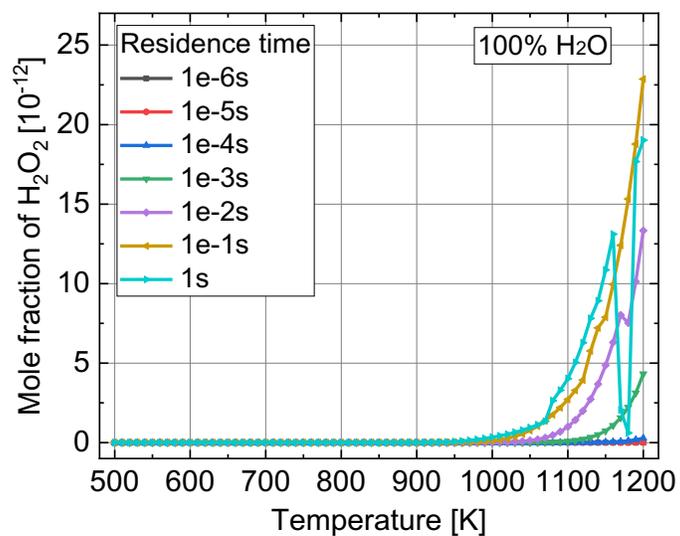

Fig. S7. Predicted $H_2O_2$ concentration at various temperatures and residence times.

From the above results, it is assumed that a significant yield of $H_2O_2$ is possible only at a high-temperature condition. However, based on the current shock wave and chemical reaction analyses, it is not possible to explain the observed formation of $H_2O_2$ in the water microdroplets.

**Section S2. Glovebox experiments in the presence of ozone – rationale and details**

The spray and condensation experiments were set up quite differently vis-à-vis the exposure to $O_3(g)$. The sprays were exposed to stable levels of ozone throughout the sample collection. This was realized by continuously supplying $O_3(g)$ to the glovebox to maintain its partial pressure. In contrast, in the condensation experiments were initiated in the presence of $O_3(g)$, but it was no longer supplied during sample collection; so, $O_3(g)$ depleted over time (Fig S8). The reasons behind the different methodologies are explained below.

Our spray setup was constantly injecting nitrogen into the glovebox at a 2.3 L/min flow rate. This nitrogen flowrate from would rapidly dilute the $O_3(g)$ in the glovebox, if we did not continuously add more $O_3(g)$. So, the partial pressure of $O_3(g)$ in the glovebox was continuously adjusted by manually controlling the inflows of ozone, pure nitrogen and clean air. For instance, if the ozone level was increasing above a certain range (Fig S8C), the nitrogen (not the shearing gas of the spray) and air flows were increased for dilution. Thus, the glovebox inlet flows were: (i) ozone (+ air), (ii) nitrogen inflow from the spray, (iii) nitrogen for dilution, and (iv) air for dilution; and there was an outlet one-way valve to let gases escape. The ozone generator, rated for 24 g-$O_3$/h production, was placed inside a sealed container with only an air inlet and air + ozone outlet. Since this setup generated an excessive ozone concentration in the outflow, way above the ppb levels we needed in this study, the air + ozone flow was split into two flows – one that was fed into the glovebox and another that was discarded to the fume hood. Additionally, since the dilutant air and nitrogen flows had significantly low relative humilities, the final RH in the glovebox varied in the range of 30%–70%.

Unlike the sprays, the condensation experiments, required relatively higher humidity to aid condensation on the cooled silicon wafers within a reasonable time. However, the dilutant air and nitrogen flows (used in the spray experiments) had a much lower humidity than required for condensation. Consequently, it was impractical to simultaneously control ozone and humidity levels in the glovebox by continuously adjusting all the gas inflows (mentioned above) and the heating rates of the water beaker or the ultrasonic humidifier power. Therefore, for the condensation experiments, we adjusted the concentration of ozone in the glovebox, then we closed all inflows of gases and placed the silicon wafers on top of ice bags to begin the condensation; the ozone concentration was then measured over the total period of condensation. It took about 40 minutes to collect adequate amount of condensate generated from the vapor supplied by humidifier or heated water (40 °C) for the HPAK analysis for $H_2O_2$. During this period, the $O_3(g)$ concentration in the gas phase gradually decreased; in some cases, to levels below the detection limit (Fig. S8A-B).

For the condensation experiments, we present the final $H_2O_2$ concentration in the condensates against the initial $O_3(g)$ concentration in Fig. 5 (green squares and red circles); while for the spray experiments Fig. 5 (blue triangles), we present the final $H_2O_2$ concentration in the collected sample against the mean $O_3(g)$ concentration during the 5 minutes of spray collection. Here, it must be recognized that the system never attained thermodynamic equilibrium, nor was it intended to. Additionally, a number of factors influenced the fate of $O_3(g)$, for instance, it reacted with and/or adsorbed onto surfaces inside of the glovebox such as the frame, nuts/bolts, electrical outlets, etc.; some $O_3$ also leaked through the outlet of the glovebox.

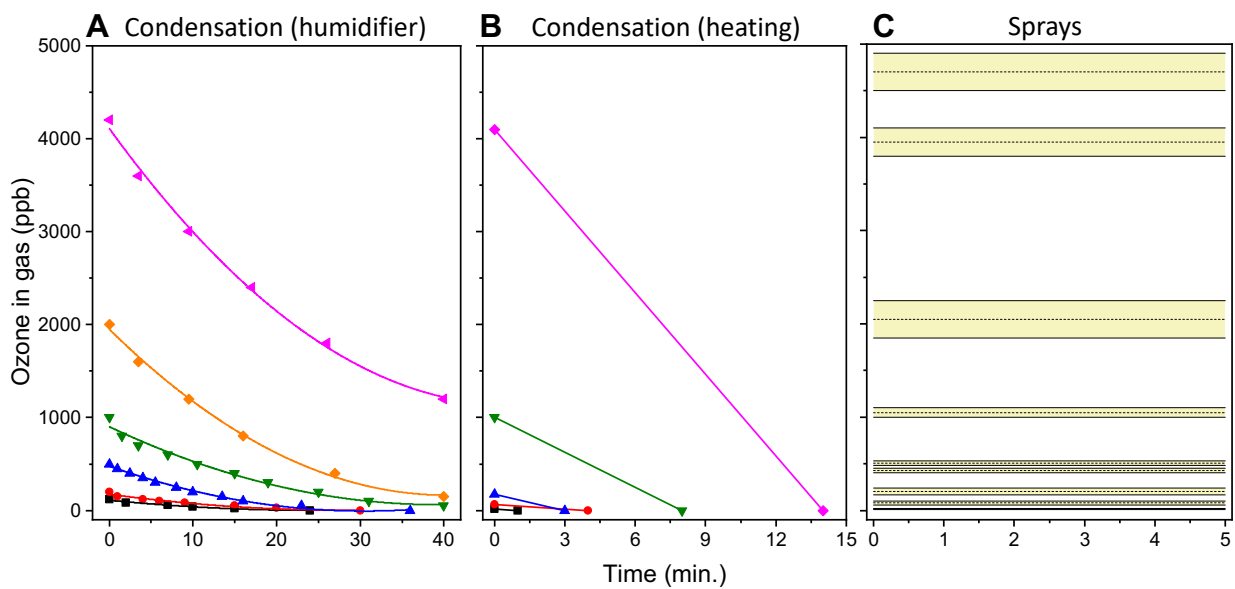

Fig. S8 – Gradual depletion of ozone from the glovebox for different initial ozone concentrations in (A) condensation by using the ultrasonic humidifier as the humidity source, and (B) condensation by heating water as the humidity source. (C) Range of ozone concentration for sprays showing no depletion. Ozone was added to the specified initial concentration in (A-B) while ozone was manually adjusted within the yellow bands (C) for the spray experiment.

Reaction with water combined with oxidation of elements inside of the glovebox (i.e., metal support, electrical outlet, etc.) and some leakage through the glovebox outlet contributed to the ozone depletion observed in Fig. S8A-B. We compared those effects by measuring depletion of ozone from the glovebox with and without water (Fig. S9); 120 mL of water was placed inside shallow containers with a surface area of 390 cm². We observed that the $O_3(g)$ depletion in the presence of water was faster. From the difference in the ozone concentration at ~55 min, and with the glovebox volume of 140 L, we estimated that ~0.7 µmol of ozone diffused into the water; assuming a 1:1 ozone to $H_2O_2$ molar conversion, the concentration of $H_2O_2$ in the liquid was expected to be ~5.8 µM. From our experimental measurements, we obtained 0.85 µM, which is ~15% of the estimated value. Despite the difference, this indicates that ozone is indeed being converted to $H_2O_2$. The difference from the measured to the calculated values could be attributed to slightly different rates of ozone depletion in the glovebox, i.e., oxidation of its internal components, or to lower ozone to peroxide conversion ratios.

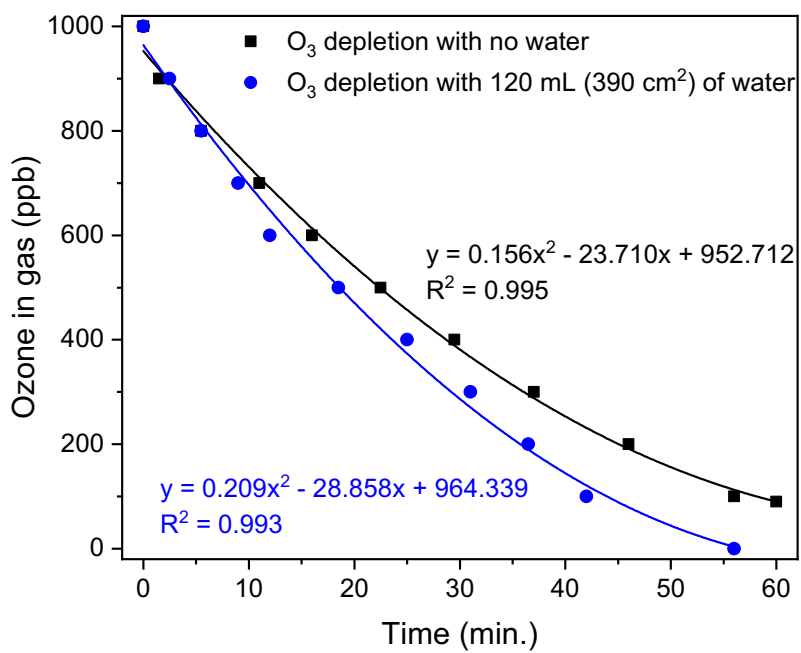

Fig. S9 – Gradual depletion of ozone from the glovebox with and without the presence of water at the same conditions.